# Molecular Beam Epitaxy Growth of Scandium Nitride on Hexagonal SiC, GaN, and AlN

Joseph Casamento[1], John Wright[1], Reet Chaudhuri[2], Huili Grace Xing[1,2], and Debdeep Jena[1,2]
[1]Department of Materials Science and Engineering, Cornell University, Ithaca, New York 14853, USA
[2]School of Electrical and Computer Engineering, Cornell University, Ithaca, New York 14853, USA

RF plasma assisted MBE growth of Scandium Nitride (ScN) thin films on GaN (0001)/SiC, AlN (0001)/$Al_2O_3$ and on 6H-SiC (0001) hexagonal substrates is found to lead to a face centered cubic (rock-salt) crystal structure with (111) out-of-plane orientation instead of hexagonal orientation. For the first time, cubic (111) twinned patterns in ScN are observed by in-situ electron diffraction during epitaxy, and the twin domains in ScN are detected by electron backscattered diffraction, and further corroborated with X-ray diffraction. The epitaxial ScN films display very smooth, sub nanometer surface roughness at a growth temperature of 750C. Temperature-dependent Hall-effect measurements indicate a constant high n-type carrier concentration of ~$1x10^{20}$/$cm^3$ and electron mobilities of ~ 20 $cm^2$/Vs.

The III-Nitride semiconductors GaN, InN, and AlN and their heterostructures have triggered a rapid expansion of photonic and electronic device applications into new wavelength, voltage, and frequency regimes. Bringing new and unique physical properties into this semiconductor family by new, epitaxially compatible nitride materials holds the promise to significantly expand what is possible today.[1]

The transition metal Sc is stable in a hexagonal crystal structure in its elemental form. The Sc ion, as a group III element, has a stable oxidation state of +3. This allows the formation of rock-salt ScN with N in the -3 oxidation state, and isovalent alloying when replacing the group III elements In, Ga, Al, or B for the III-nitride materials family. This feature, combined with predictions of a metastable nonpolar h-BN-like, and wurtzite crystal structures [2,3], has led to significant interest in ScN and its alloys. These alloys promise to bring to the established GaN and AlN based electronics and photonics family, missing properties such as plasmonic [4], thermoelectric [5,6], extremely high piezoelectric [7,8], and also ferroelectric behavior in ScAlN [9-11].

As Sc-based alloys with AlN and GaN (such as ferroelectric ScAlN) are beginning to be explored by epitaxy [12,13], it is essential that the epitaxial growth of the limiting binary ScN thin films, and their physical properties be understood. The binary compound Scandium Nitride (ScN) is part of the family of the transition metal nitride semiconductors, with desirable physical properties such as high hardness, mechanical strength, and high temperature stability [14-17]. Its equilibrium phase has a face-centered cubic rocksalt (NaCl) structure with a lattice constant of 4.505 angstroms [18]. The (111) lattice constant of cubic ScN is nearly lattice-matched (only ~0.1% difference) to the c-plane lattice constant of wurtzite GaN [19]. This has led to avenues for the use of ScN as a dislocation reduction buffer layer, and for in-situ ohmic contacts for GaN based devices [20]. Earlier studies have reported the growth of binary ScN thin films by reactive magnetron sputtering, hydride vapor phase epitaxy, and gas-source and plasma MBE on Si, $Al_2O_3$, SiC, and MgO surfaces [21-29]. Very few reports exist of ScN epitaxy on GaN [30], and no results have yet been shown for growth on AlN. One literature report mentions mixed (110) and (111) orientations for ScN grown on AlN, but the data is not presented [31]. Here, we present a comparative study of the plasma-MBE growth, structure, surface morphology, and electrical transport properties of ScN thin films (~ 30 nm) on c-plane GaN, AlN, and SiC surfaces. We find that extremely smooth epitaxial thin films of high crystalline quality of rocksalt ScN grow with their 111 axes aligned with the polar 0001 c-axis of the wurtzite substrates of GaN, AlN, and SiC, and are unintentionally n-type doped with degenerate electron concentrations.

ScN films were deposited on various substrates using a Veeco® GenXplor MBE system in which the idle-state base pressure is $5 \times 10^{-10}$ torr. Solid Sc source of 99.99 % purity on a rare earth element basis from American Elements was evaporated using a Telemark® electron beam evaporation system in the MBE environment. An electron beam is steered into a W crucible with a magnetic coil to create the Sc flux. Flux stability was achieved with an Inficon® electron impact emission spectroscopy (EIES) system by directly measuring the Sc atomic optical emission spectra. Nitrogen was supplied using a Veeco® RF UNI-Bulb plasma source, with growth pressure of approximately $10^{-5}$ torr. In-situ monitoring of film growth was performed using a KSA

Instruments reflection high-energy electron diffraction (RHEED) apparatus with a Staib electron gun operating at 15 kV and 1.5 A. After epitaxial growth, the film thickness and orientation were characterized using a PanAlytical X'Pert Pro XRD setup at 45 kV, 40 mA with Cu Kα radiation (1.5406 Å). Grain topography and orientation was investigated using an Electron Back-Scatter Diffraction (EBSD) detector with a 70-degree geometric tilt correction in a Tescan Mira SEM system with an operating pressure of $10^{-5}$ torr. Kikuchi patterns were indexed to face centered cubic ScN (Fm-3m space group, 225). Temperature-dependent Hall-effect measurements were taken using indium contacts in a Van der Pauw geometry in a Lakeshore® system with a 1T magnet from room temperature to 20K.

ScN films were grown in nitrogen-rich conditions at a thermocouple temperature of 750C, a Sc flux of 0.16 Å/s, a Nitrogen plasma condition of 1.95 standard cubic centimeter per minute (sccm) and 200W, and a chamber pressure of $1.5 \times 10^{-5}$ torr. Nitrogen rich growth conditions were utilized to suppress nitrogen vacancy formation and maintain a 1:1 Sc to N stoichiometry in the epitaxial film [32]. ScN layer thicknesses were 30 nm with a growth rate of 30 nm/hour.

During growth, RHEED was utilized to continuously monitor the surface crystal structure of ScN. During growth on hexagonal substrates, the (110) azimuth evolved from first order 1x1 streaks for GaN to pairs of symmetric spots on either side of the original first order streaks for ScN. These spots can be viewed as rotated variants of two separate overlaid (1-10) zone axes, indexed as pairs of (111) and (002) families of planes, as shown in Figure 1. This RHEED pattern and indexing has not been reported before for ScN (111) films but was seen in Cu (111) thin films grown on $Al_2O_3$[33].

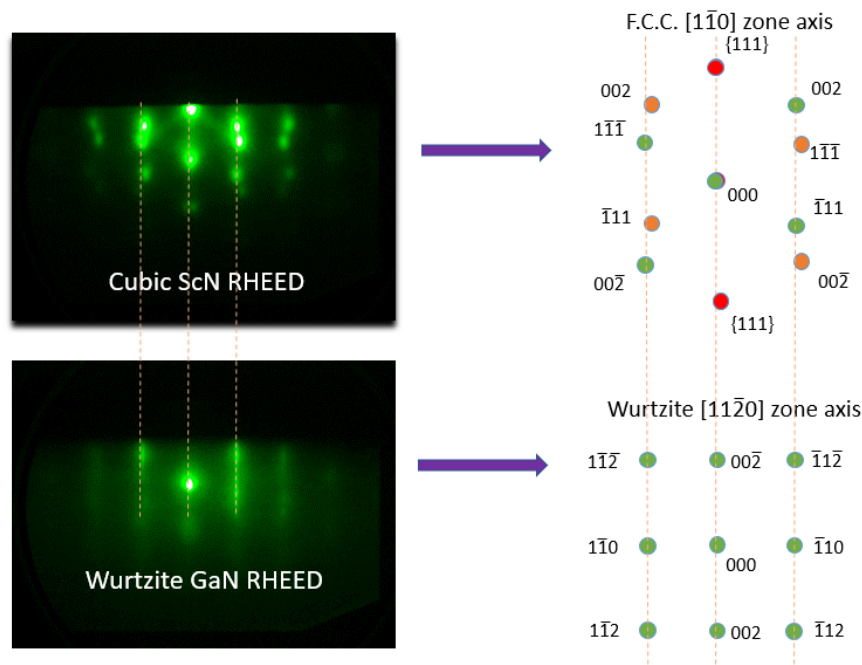

Figure 1: RHEED pattern during evolution along the [110] azimuth from GaN (0001) to ScN (111). 111 and 002 pairs of kinematically allowed diffraction spots are symmetrically rotated about the zone center, illustrating ScN grows as a cubic-twinned crystal.

The observed RHEED pattern implies the existence of twin domains that are expected to result from the symmetry constraints encountered upon growing a three-fold symmetric cubic crystal on a six-fold symmetric hexagonal substrate. The (110) zone axis for a cubic single crystal would only have two 111 and two 002 diffraction spots, one each on each side of the zone center (000). Here, four 111 and four 002 diffraction spots are seen, two on each side of the zone center, indicating a (111) cubic twinned crystal.

To further test for the existence of twin domains in the epitaxial ScN layers, electron back-scatter diffraction (EBSD) measurements were performed on ScN films grown on 6H SiC and GaN/SiC template substrate (Figure 2). In EBSD, the incident beam-sample-detector geometry is such that backscattered electrons escape the sample through the Bragg angle and diffract and form Kikuchi bands. If the chemical composition is known, the orientation sensitive Kikuchi bands can be indexed to determine the crystal orientation in different regions. Accordingly, Kikuchi patterns were indexed to face centered cubic (FCC) ScN to verify the symmetry of the ScN film. EBSD topography images showed striped patterns with alternating orientations. Two different colors indicate two different orientations, as expected for domains on either side of a twin boundary. Pole figures with the (111) direction out of plane indicated six-fold symmetry, as shown in Figure 2 for samples grown on 6H-SiC and GaN/SiC. The corresponding AFM images indicate highly smooth surface morphologies, with rms roughness below 1 nm for 10x10 micron scans for ScN layers that are 30 nm thick. Grain misorientation statistics (not shown) indicated the ScN epitaxial film was entirely (111) oriented, with 60 degrees being the dominant in-plane misorientation angle.

A.)

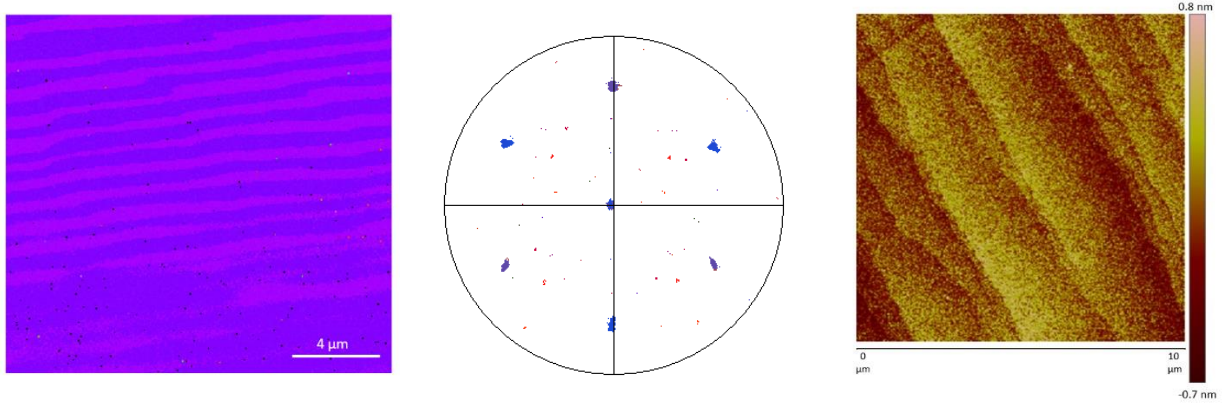

B.)

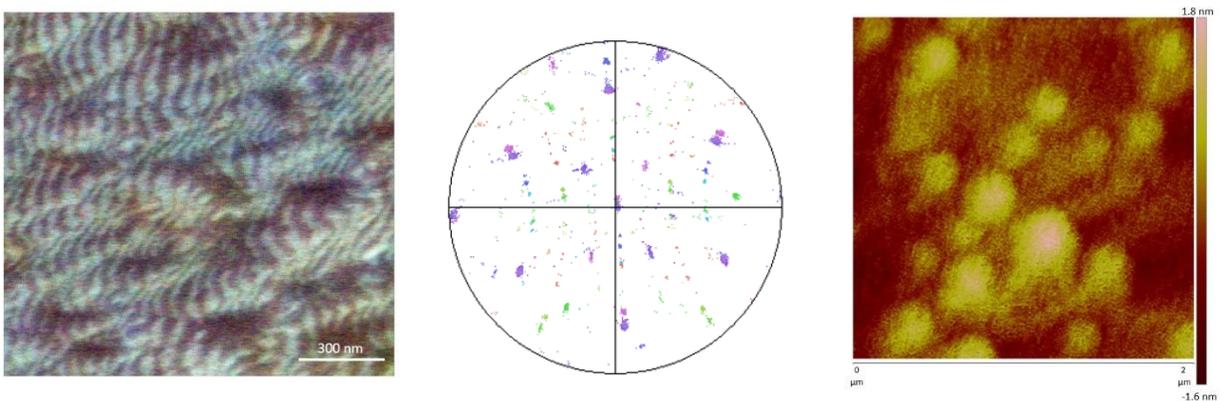

Figure 2 A.) EBSD topography image of ScN grown on 6H-SiC (left). EBSD pole figure of ScN with {111} direction out of plane (center). 10x10μm AFM image(right). B.) EBSD topography of ScN grown on GaN/SiC (left). EBSD pole figure of ScN with {111} direction out of plane (center). 2x2μm AFM image (right). Alternating stripe patterns in the EBSD topography images indicate domains with 180˚ orientation device, with six spots in the EBSD pole figures indicating six-fold symmetry in the {111} direction.

Hence, we find that the ScN samples are not polycrystalline with random grain orientations, but rather highly oriented crystals with grain misorientation. EBSD determined twin domain sizes differed from approximately 800 nm for samples grown on SiC and 60 nm when grown on GaN/SiC. It is currently unclear if step-terraces or other surface features from the substrate such as threading dislocations influence the nucleation of cubic-twin domains and subsequent size of these domains. Growth on bulk substrates combined with transmission electron microscopy (TEM) studies can answer this question definitively in the future.

The orientation of the ScN crystal planes was further assessed using X-Ray diffraction (XRD) measurements, as seen in Figure 3. Whereas the ScN peak is clearly resolved for growths on SiC and AlN, the films grown on GaN/SiC template substrates show no separate ScN peak. This is because the ScN peak lies within the background of the GaN (002) peak. This is expected given the lattice constant of ScN (111) is only 0.18% different from the in-plane lattice constant of wurtzite GaN, 3.189Å. For films grown on AlN/Al$_2$O$_3$ and SiC, separate ScN (111) peaks are observed, yet none of the samples showed cubic ScN (00l) reflections, indicating ScN forms a highly oriented film with the (111) direction aligned along the 0001 c-axis of the substrate and perpendicular to the growth plane. The peak positions of ScN (111) in the films grown on SiC and are consistent with those reported in literature occurring near $2\theta = 34.5°$ [25]. Scans of the ScN (224) peak for a sample grown on SiC indicate six-fold symmetry, with peaks separated by sixty degrees. This is further support for the nature of a cubic-twinned ScN (111) film.

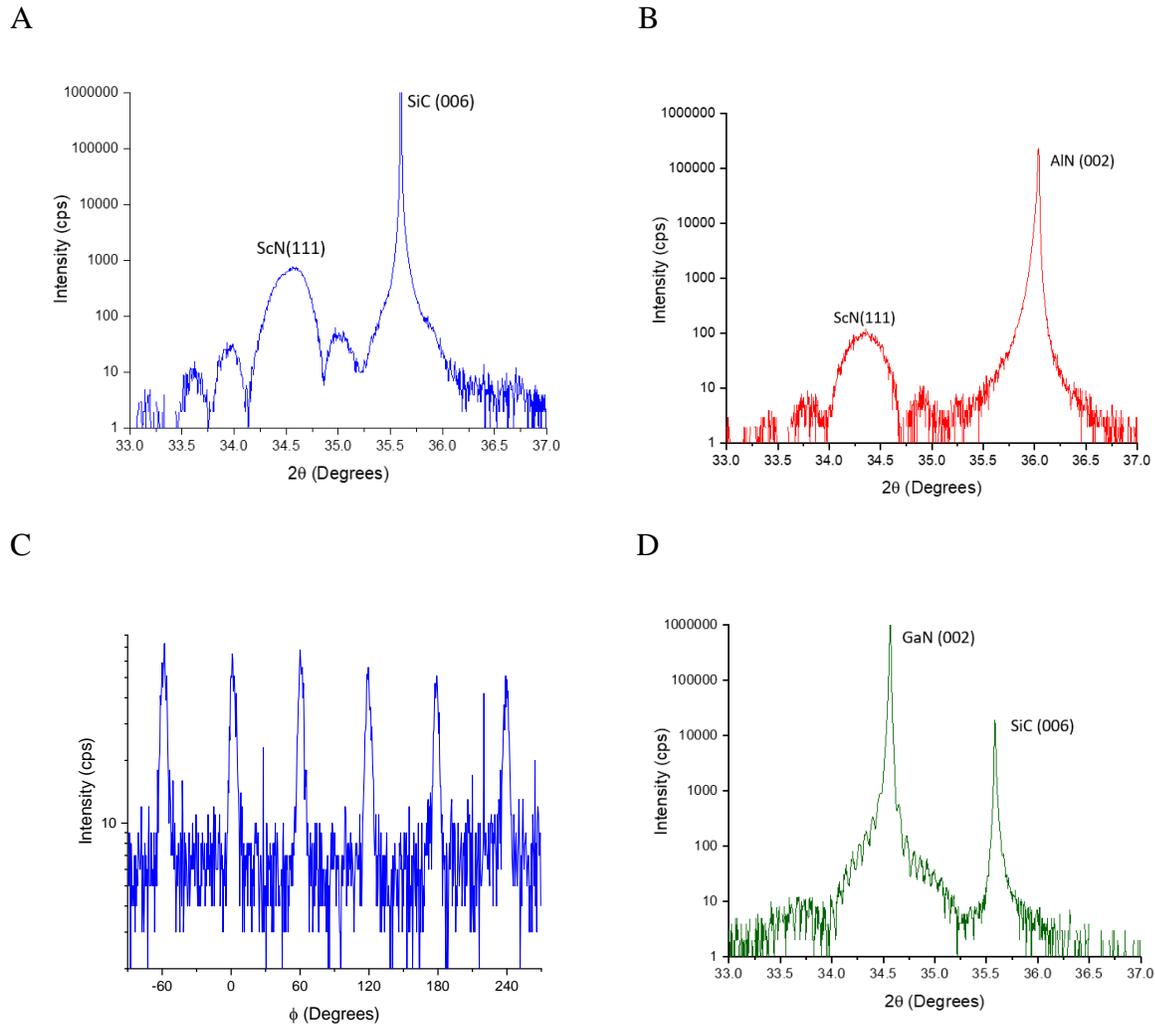

Figure 3: A-B) XRD 2theta-omega scans of ScN grown on 6H-SiC, AlN/Al$_2$O$_3$. C-D) XRD Phi Scan of ScN (224) peak grown on 6H-SiC, XRD 2theta-omega of ScN grown on GaN/SiC. The XRD results indicate ScN grows epitaxially in an (111) orientation on the respective substrates, with six-fold in-plane rotational symmetry as seen from the Phi scans.

The ScN films grown on all three hexagonal substrates (SiC, AlN, GaN) adopted a twinned face centered cubic crystal structure instead of a hexagonal structure. Reasons for the stability of the cubic phase of ScN can be found in the principles of molecular orbital energies. For binary nitrides, the three 2p orbitals of Nitrogen anion and the three lower $t_{2g}$ states (of the crystal-field split five 3d orbitals) of the Scandium transition metal cation hybridize to form bonding and antibonding orbitals. The remaining two 3d orbitals remain as nonbonding energy states, with energy levels close to the transition metal 3d orbitals. A schematic of this *pd* coupling dominated bonding is shown in Figure 4 and a similar bonding diagram has been illustrated previously in the literature.[15]

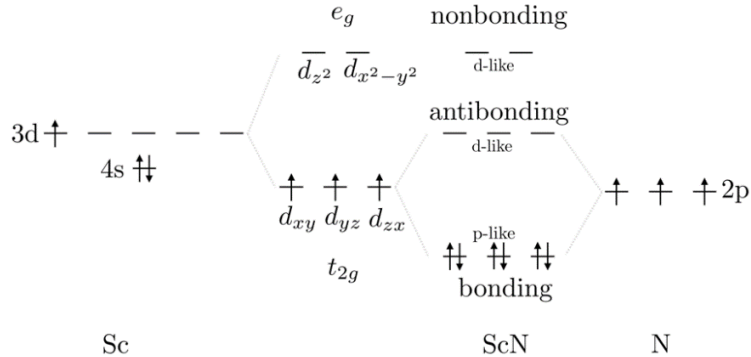

Figure 4: Bonding schematic of rocksalt ScN. Octahedral coordination causes d-orbital crystal field splitting into $t_{2g}$ and $e_g$ orbitals. All electrons occupy bonding states, indicating an extremely stable cubic structure.

The three outermost electrons of Sc ([Ar]$4s^2 3d^1$) bond with three N 2p electrons in a crystal to make $Sc^{+3}$. Adding more electrons (e.g using Ti, V, Cr, Mn… instead of Sc) will populate the antibonding and then nonbonding energy states, making the rock-salt crystal structure less stable, and the crystal metallic. On the other hand, ScN is an extremely stable rock-salt semiconductor because of the filled bonding states, with a completely empty band in the $Sc^{+3}$ configuration. ScN is a semiconductor with an indirect bandgap of 0.9 eV and a direct bandgap of ~2.1 eV [34]. XPS results (not shown) verify the presence of Sc-N bonding and optical absorption measurements (not shown) show a weak band edge absorption near 2.1 eV.

The electrical transport properties of the 30 nm thick MBE grown ScN on GaN/SiC and AlN/$Al_2O_3$ template substrates was assessed using temperature-dependent Hall-effect measurements from room temperature to 20K, as shown in Figure 5. A high n-type carrier concentration independent of temperature was observed. The lack of carrier freezeout at low temperatures indicates the nominally undoped ScN obtained in this work is a degenerately doped semiconductor. As shown in Figure 5, the carrier concentrations and mobilities were $1.55 \times 10^{20}/cm^3$ and 23 $cm^2/Vs$ and $1.05 \times 10^{20}/cm^3$ and 11 $cm^2/Vs$ for samples grown on GaN/SiC and AlN/$Al_2O_3$, respectively. The high carrier concentration is an order of magnitude higher than reported values of the conduction band density of states for many semiconductors, giving support for its degenerate, metallic electrical behavior. At this stage it is unclear if ScN is strained or relaxed to AlN and if the crystal quality and carrier mobility is affected by the approximately 2.4% lattice mismatch. The obtained mobility values are lower than those reported previously [6,27-29,31,34,36,38] for similar carrier concentrations and this may be due to increased impurity scattering from domain boundaries in (111) oriented films. The (111) surface is a higher energy surface than the (100) surface for a face centered cubic crystal. Higher surface energy planes are more likely to trap impurities like oxygen, potentially due to a decrease in adatom mobility. This has been shown in ScN, where oxygen concentration was higher for cubic-twinned (111) growth on c-plane sapphire compared to untwinned cubic (111) growth on MgO (111) [26]. This has also been reported in the case of Si and Ti segregation to twin domain boundaries in $MgAl_2O_4$ [35].

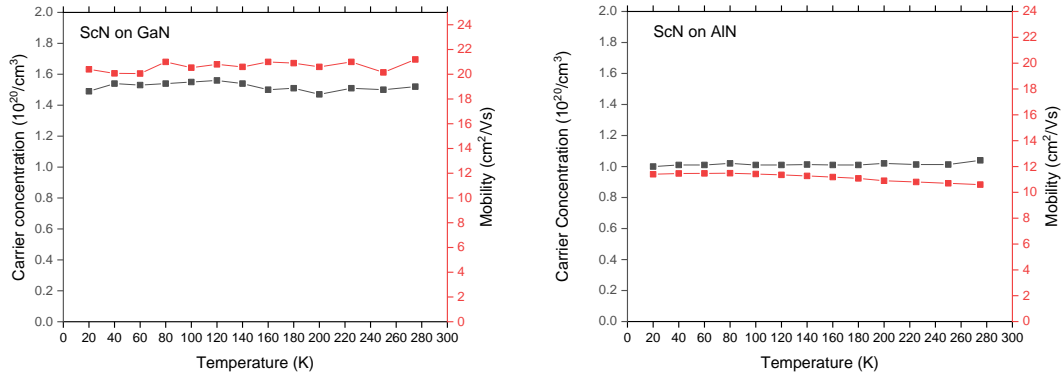

Figure 5: Temperature dependent hall data of ScN grown on GaN/SiC (left) and $AlN/Al_2O_3$ (right). Carrier concentrations and mobilities whose magnitude does not change indicates degenerate doping behavior.

A high electron carrier concentration has previously been reported in nominally undoped ScN, with possible causes being linked to nitrogen vacancies, Sc-N antisite defects, and atomic level concentrations of oxygen and fluorine originating from source and crucible material. Density functional theory (DFT) calculations have shown oxygen substitutional defects are lower in formation energy than the other defect mechanisms mentioned above [36]. Support for oxygen incorporating in the film and donating electrons comes from Scandium metal's high affinity for oxygen, as evidenced in its large negative enthalpy of formation for $Sc_2O_3$ from Ellingham diagrams [37]. Carrier mobilities up to 100 $cm^2$/Vs at carrier concentrations of $10^{21}$/$cm^3$ have been previously obtained in ScN on MgO (001) [38] and mobilities up to 284 $cm^2$/Vs at a carrier concentration of $3.7\times10^{18}$/$cm^3$ have been obtained in ScN grown using hydride vapor phase epitaxy (HVPE) on m and r-plane sapphire substrates. The lower carrier concentration and higher mobility in HVPE growth was due to a reduction in impurities in the film, notably oxygen concentration, through utilization of 6N (99.9999%) pure $ScCl_3$ and $NH_3$ as the source materials instead of Sc metal [31]. This points towards an important role that impurities have on determining the carrier concentration and limiting the mobility of ScN films. Using ionized impurity scattering models [39], assuming all donors are ionized and a density of states effective mass of ~0.35$m_e$ where $m_e$ is the free electron mass, a close agreement to the obtained mobility values is found. However, we point out that the interface between rock-salt ScN and hexagonal GaN or AlN is a nonpolar/polar interface, and the polar discontinuity across the interface, assisted by the conduction and valence band offsets can give rise to mobile carrier concentrations even in the absence of defects and impurities [40]. Similar polar/non-polar interfaces have been found in several oxides, but not in the nitride crystals yet. Future work involving the measurement of electronic structure and electron mobility will give further insight into the mobility limiting mechanisms in ScN, and more importantly, the origin of the mobile charges in the bulk and at interfaces.

In this work, we have reported the MBE growth of highly crystalline ScN thin films on hexagonal GaN, AlN, and SiC substrates. ScN films exhibited solely cubic twinned (111) orientation on all three hexagonal substrates, and did not adopt a hexagonal crystal structure. ScN films exhibited large n-type carrier concentrations of approximately $10^{20}$/cm$^3$ with mobilities of approximately 20 cm$^2$/Vs. For the first time, SEM-EBSD and in-situ RHEED patterns confirm cubic-twinned domains and grain orientation in the epitaxial ScN grown on the hexagonal wide-bandgap GaN, SiC, and AlN surfaces. This work sheds light on the fundamental cubic stability of ScN and provides a roadmap for future work regarding the analysis of ScN growth thermodynamics, epitaxial stabilization, and integration in novel III-nitride device architectures. The findings of the limiting case of ScN growth on AlN and ScN should be a valuable guide towards the future investigation of highly piezoelectric and ferroelectric GaScN and ScAlN alloys.

This work made use of the Cornell Center for Materials Research Shared Facilities which are supported through the NSF MRSEC program (DMR-1719875). This work was also performed in part at the Cornell NanoScale Facility, a member of the National Nanotechnology Coordinated Infrastructure (NNCI), which is supported by the National Science Foundation (Grant NNCI-1542081). The authors would like to acknowledge the Air Force Office of Scientific Research, Grant AFSOR FA9550-17-1-0048, for the support of this work. The authors would also like to acknowledge Don Werder for discussions related to EBSD and SEM and Mike Hawkridge for discussions related to XRD.